\documentclass[12pt]{article}

\usepackage{cite}

\def\eptwo{\left\{ \phantom{|}^{\mu\nu}_{ab} \right\}}
\def\epthree{\left\{ \phantom{|}^{\mu\nu\alpha}_{abc} \right\}}
\def\epfour{\left\{ \phantom{|}^{\mu\nu\alpha\beta}_{abcd} \right\}}

\author{Yu.~M.~Zinoviev
       \thanks{E-mail address: Yurii.Zinoviev@ihep.ru} \\
        {\it Institute for High Energy Physics} \\
        {\it of National Research Center "Kurchatov Institute"} \\
        {\it Protvino, Moscow Region, 142280, Russia}}
\title{Towards the Fradkin-Vasiliev formalism \\ in three dimensions}

\date{}

\begin{document}

\maketitle

\begin{abstract}
In this paper we show that using frame-like gauge invariant
formulation for the massive bosonic and fermionic fields in three
dimensions the free Lagrangians for these fields can be rewritten in
the explicitly gauge invariant form in terms of the appropriately
chosen set of gauge invariant objects. This in turn opens the
possibility to apply the Fradkin-Vasiliev formalism to the
investigation of possible interactions of such fields.
\end{abstract}

\thispagestyle{empty}
\newpage
\setcounter{page}{1}

\section*{Introduction}

One the effective ways to investigate interactions for higher spin
fields in the Fradkin-Vasiliev formalism \cite{FV87,FV87a} (see also
\cite{Vas11,BPS12}). Initially this formalism was developed for the
massless higher spin fields (some examples may be found in 
\cite{Vas01,AV02,Alk10,BSZ11,BS11}). But the most important
ingredients of such formalism are frame-like formalism and gauge
invariance and the frame-like gauge invariant description exists for
the massive higher spins as well \cite{Zin08b,PV10}. Thus, 
in-principle, this formalism can be applied to the investigation of
possible interactions for any system with massive and/or (partially)
massless fields (some examples can be found in
\cite{Zin10a,Zin11,Zin14,BSZ14}).

At the same time it is a common belief that the Fradkin-Vasiliev
formalism operates in dimensions equal or greater then four only.
Indeed, as far as the massless higher spins in $d=3$ are concerned,
one has mostly deals with the Chern-Simons theories (e.g.
\cite{Ble89,AT86,CFPT10,Zin14a}) that is tightly connected with the
fact that such massless fields do not have any local physical degrees
of freedom being pure gauges. But there are three cases when higher
spin fields in $d=3$ do have some physical degrees of freedom: bosonic
massive field, bosonic partially massless field of the maximal depth
and fermionic massive one. In this paper we show that for all these
cases one can rewrite the free Lagrangians in the explicitly gauge
invariant way in terms of the appropriate set of gauge invariant
objects. This, in turn, can serves as a starting point for the
application of Fradkin-Vasiliev formalism to investigation of possible
interactions for such fields. Our construction will be based on the
frame-like gauge invariant formulation for the massive bosonic and
fermionic higher spin fields in $d=3$ \cite{BSZ12a,BSZ14a} (see also
\cite{BSZ12b,BSZ15,BSZ16}).

The paper is organized as follows. In Section 1 we give a short review
of the main features of the Fradkin-Vasiliev formalism. We separately
consider massless and massive case and show that in the massive case
there is a possibility to extend the formalism to three dimensions. In
Section 2 we consider bosonic higher spin field (both massive as well
as partially massless cases) while in Section 3 we consider massive
fermionic case. In the Appendix using massive spin 2 as the simplest
non-trivial example we show that it is indeed possible to adopt the
results of \cite{PV10} to three dimensions. But the analogous result
for the arbitrary spin would require a lot of rather complicated
calculations so in the main part of the paper we find all necessary
formulas from scratch directly in $d=3$.

\noindent
{\bf Notations and conventions.} We use a frame-like multispinor
formalism where all objects (one-forms or zero-forms) have local
indices which are completely symmetric spinor ones. To simplify
expressions we will use condensed notations for the spinor indices
such that e.g.
$$
\Omega^{\alpha(2k)} = \Omega^{(\alpha_1\alpha_2 \dots \alpha_{2k})}
$$
Also we will always assume that spinor indices denoted by the same
letter and placed on the same level are symmetrized, e.g.
$$
\Omega^{\alpha(2k)} \zeta^\alpha = \Omega^{(\alpha_1\dots \alpha_{2k}}
\zeta^{\alpha_{2k+1})}
$$
$AdS_3$ space will be described by the background frame (one-form)
$e^{\alpha(2)}$ and the covariant derivative $D$ normalized so that
$$
D \wedge D \zeta^\alpha = - \lambda^2 E^\alpha{}_\beta \zeta^\beta
$$
where two-form $E^{\alpha(2)}$ is defined as follows:
$$
e^{\alpha(2)} \wedge e^{\beta(2)} = \varepsilon^{\alpha\beta}
E^{\alpha\beta}
$$
In the most part of the paper the wedge product sign $\wedge$ will be
omitted.

\section{Fradkin-Vasiliev formalism}

In this section we provide brief review of the Fradkin-Vasiliev
formalism. Our aim here to look for the possibilities to apply this
formalism to higher spins in three dimensions. We begin with the
massless case and then we consider the massive one.

\subsection{Massless case}

In the frame-like formalism the description of the massless higher
spin requires a collection of one-forms (physical, auxiliary and extra
ones) with some local indices which we collectively denote as $\Phi$.
Each field is a gauge field with its own gauge transformation
(schematically):
$$
\delta \Phi \sim D \xi \oplus e \xi
$$
For each field (physical, auxiliary or extra one) we can construct a
gauge invariant two-form (curvature):
$$
{\cal R} \sim D \wedge \Phi \oplus e \wedge \Phi
$$
The free Lagrangian can then be rewritten in terms of these curvatures
in the explicitly gauge invariant form:
$$
{\cal L}_0 = \sum c_k {\cal R}_k \wedge {\cal R}_k
$$
where coefficients are fixed by the so called extra field decoupling
condition. As it well known for the massless fields such description
requires that the cosmological constant be non zero. At the same time
from the structure of the Lagrangian (the sum of the squares of 
two-forms) it is clear that the space-time dimension must be greater
or equal to four. Thus it is not possible to apply such formalism to
the massless higher spins in $d=3$.

Let us turn to the construction of cubic vertices in this formalism.
The first step is to construct the most general quadratic deformations
for all these gauge invariant curvatures:
$$
{\cal R} \Rightarrow \hat{\cal R} = {\cal R} \oplus \Phi \wedge \Phi
$$
Here the most important requirement is that these deformed curvatures
transform covariantly under the all gauge symmetries:
$$
\delta \hat{\cal R} \sim {\cal R} \xi
$$
Then one consider the following ansatz for the interacting Lagrangian:
$$
{\cal L} \sim \sum \hat{\cal R} \wedge \hat{\cal R} \oplus \sum
{\cal R} \wedge {\cal R} \wedge \Phi
$$
Here the first part is just the free Lagrangian but with the
curvatures replaced by the deformed ones, while the second part
contains all possible abelian (or Chern-Simons like) vertices. In the
linear approximation the variations of this Lagrangian lead to the
expressions quadratic in the curvatures and the requirement for the
Lagrangian be gauge invariant is reduced to a set of simple algebraic
equations on the coefficients. It has been shown \cite{Vas11} that all
non-trivial cubic vertices for the massless higher spins can be
reproduced in such a way.

\subsection{Massive case}

As it can be seen from the massless case two most important
ingredients of the Fradkin-Vasiliev formalism are frame-like formalism
and gauge invariance. But the frame-like gauge invariant description
exists for the massive higher spins as well \cite{Zin08b,PV10} and it
opens the possibility to apply this formalism to the construction of
the cubic vertices containing massive and/or massless higher spins.
Let us stress the main differences between massless and massive
cases.

The frame-like gauge invariant formulations necessarily requires not
only a set of one-forms $\Phi$ but also a set of zero-forms $C$ which
transform non-trivially under the gauge transformations:
$$
\delta \Phi \sim D\xi + \dots, \qquad \delta C \sim \xi
$$
Thus the zero-forms play the role of the Stueckelberg fields and their
appearance is quite natural because in the massive case we have to
expect that all gauge symmetries must be spontaneously broken.

Similarly to the massless case, for each field one can also construct
a gauge invariant object (two-form for the one-form field and one-form
for the zero-form one):
\begin{eqnarray*}
{\cal R} &\sim& D \Phi \oplus e \Phi \oplus eeC \\
{\cal B} &\sim& D C \oplus \Phi \oplus e C
\end{eqnarray*}

Using these gauge invariant objects one can rewrite the free
Lagrangian in the explicitly gauge invariant form:
$$
{\cal L}_0 \sim {\cal R} {\cal R} \oplus {\cal R} {\cal B} \oplus
{\cal B} {\cal B}
$$
The coefficients in these expression also must satisfy the extra field
decoupling condition, but in the massive case  their solution is not
unique. The reason is that some combinations of these quadratic terms
form total derivative. All such combinations can be systematically
generated using differential identities on these gauge invariant
objects. Now if we managed to find a solution where all the terms of
the type ${\cal R} {\cal R}$ are absent we will obtain the form of the
free Lagrangian that is valid in three dimensions as well. In the
Appendix, using the massive spin-2 case as the simplest non-trivial
example, we show that it is indeed possible. But for the arbitrary
spin case it would require a lot of complicated calculations. So in
this paper we will work directly in three dimensions and we will
obtain all necessary formulas from scratch. Note also, that contrary
to the massless case here the cosmological constant need not be
non-zero so that such formalism works in the flat Minkowski space as
well.

Let us turn to the construction of cubic vertices. Here we also begin
with the most general quadratic deformations for all gauge invariant
objects:
\begin{eqnarray*}
\hat{\cal R} &\sim& {\cal R} \oplus \Phi \Phi \oplus e \Phi C \oplus e
e C C \\
\hat{\cal B} &\sim& {\cal B} \oplus \Phi C \oplus e C C
\end{eqnarray*}

and require that they transform covariantly:
$$
\delta \hat{\cal R} \sim {\cal R} \xi, \qquad
\delta \hat{\cal B} \sim {\cal B} \xi
$$
Due to the presence of the zero-forms there exists a lot of possible
field re-definitions that one has take into account:
\begin{eqnarray*}
\Phi &\Rightarrow& \Phi \oplus \Phi C \oplus e C C \\
C &\Rightarrow& C \oplus C C
\end{eqnarray*}
Then the interacting Lagrangian can be constructed as the free
Lagrangian where all gauge invariant objects are replaced by the
deformed one plus all possible abelian vertices.

\section{Integer spin}

In this section we consider massive bosonic field with integer spin $s
\ge 2$.

\subsection{General massive case}

The gauge invariant description for massive bosonic spin-$s$ field
\cite{BSZ12a} uses a collection of massless fields with spins $s$,
$s-1$, $\dots$, 0. In the frame-like approach we need pairs of
one-forms ($\Omega^{\alpha(2k)}$, $\Phi^{\alpha(2k)}$), $1 \le k \le
s-1$, a zero-form $B^{\alpha(2)}$ and one-form $A$ for the spin-1
component as well as two zero-forms ($\pi^{\alpha(2)}$, $\varphi$) for
the spin-0 one. The Lagrangian describing massive spin-$s$ field in
the $(A)dS_3$ background has the form:
\begin{eqnarray}
{\cal L}_0 &=& \sum_{k=1}^{s-1} (-1)^{k+1} [ k 
\Omega_{\alpha(2k-1)\beta} e^\beta{}_\gamma 
\Omega^{\alpha(2k-1)\gamma} + \Omega_{\alpha(2k)} D \Phi^{\alpha(2k)}
 ] \nonumber \\
 && + E B_{\alpha\beta} B^{\alpha\beta} - B_{\alpha\beta} 
e^{\alpha\beta} D A - E \pi_{\alpha\beta} \pi^{\alpha\beta} +
\pi_{\alpha\beta} E^{\alpha\beta} D \varphi \nonumber \\
 && + \sum_{k=1}^{s-2} (-1)^{k+1} b_k [ - \frac{(k+2)}{k} 
\Omega_{\alpha(2)\beta(2k)} e^{\alpha(2)} \Phi^{\beta(2k)} + 
\Omega_{\alpha(2k)} e_{\beta(2)} \Phi^{\alpha(2k)\beta(2)} ] \nonumber
\\
 && + 2b_0 \Omega_{\alpha(2)} e^{\alpha(2)} A - b_0 
\Phi_{\alpha\beta} E^\beta{}_\gamma B^{\alpha\gamma} + 8c_1
\pi_{\alpha\beta} E^{\alpha\beta} A \nonumber \\
 && + \sum_{k=1}^{s-1} (-1)^{k+1} c_k \Phi_{\alpha(2k-1)\beta} 
e^\beta{}_\gamma \Phi^{\alpha(2k-1)\gamma} + \frac{Msb_0}{2} 
\Phi_{\alpha(2)} E^{\alpha(2)} \varphi + \frac{3b_0{}^2}{2}
 E \varphi^2
\end{eqnarray}
where
$$
b(k)^2 = \frac{k(s+k+1)(s-k-1)}{2(k+1)(k+2)(2k+3)} 
[ m^2 - (s+k)(s-k-2) \Lambda ]
$$
$$
b_0{}^2 = \frac{(s+1)(s-1)}{3} [ m^2 - s(s-2) \Lambda ]
$$
$$
c_k = \frac{s^2M^2}{4k(k+1)^2}, \qquad
M^2 = m^2 - (s-1)^2 \Lambda
$$
Note that the structure of this Lagrangian corresponds to the general
pattern for the gauge invariant description of massive higher spin
fields. Namely, the first two lines and the last one are just the sum
of kinetic and mass-like terms for all components, while the third and
the fourth lines contain cross-terms gluing all these components
together. The most important property of this approach (that
completely determines the very structure and all the coefficients) is
that it allows one to keep all the gauge symmetries that massless
components initially have. Indeed, this Lagrangian is invariant under
the following set of gauge transformations:
\begin{eqnarray}
\delta \Omega^{\alpha(2k)} &=& D \eta^{\alpha(2k)} + 
\frac{(k+2)b_k}{k} e_{\beta(2)} \eta^{\alpha(2k)\beta(2)} \nonumber \\
 && + \frac{b_{k-1}}{k(2k-1)} e^{\alpha(2)} \eta^{\alpha(2k-2)} 
 + \frac{c_k}{k} e^\alpha{}_\beta \xi^{\alpha(2k-1)\beta} \nonumber \\
\delta \Phi^{\alpha(2k)} &=& D \xi^{\alpha(2k)} + e^\alpha{}_\beta
\eta^{\alpha(2k-1)\beta} + b_k e_{\beta(2)} 
\xi^{\alpha(2k)\beta(2)} \nonumber \\
 && + \frac{(k+1)b_{k-1}}{k(k-1)(2k-1)} e^{\alpha(2)}
\xi^{\alpha(2k-2)} \nonumber \\
\delta \Omega^{\alpha(2)} &=& D \eta^{\alpha(2)} + 3b_1
e_{\beta(2)} \eta^{\alpha(2)\beta(2)} + c_1
e^\alpha{}_\gamma \xi^{\alpha\gamma} \\
\delta \Phi^{\alpha(2)} &=& D \xi^{\alpha(2)} + 
e^\alpha{}_\gamma \eta^{\alpha\gamma} + b_1 e_{\beta(2)}
\xi^{\alpha(2)\beta(2)}  + 2b_0 e^{\alpha(2)} \xi \nonumber \\
\delta B^{\alpha(2)} &=& 2b_0 \eta^{\alpha(2)}, \qquad
\delta A = D \xi + \frac{b_0}{4} e_{\alpha(2)} \xi^{\alpha(2)}
\nonumber \\
\delta \pi^{\alpha(2)} &=& \frac{Msb_0}{2} \xi^{\alpha(2)}, \qquad
\delta \varphi = - 2Ms \xi \nonumber
\end{eqnarray}

As is now very well known, in the de Sitter space ($\Lambda > 0$)
there exist a number of special values for the mass $m$ when one of
the parameters $b(k_0) = 0$. In this case the whole Lagrangian
decomposes into two independent subsystems one of which describes a
so-called partially massless spin-$s$ field. But most of these
partially massless fields in $d=3$ do not have any physical degrees of
freedom (similarly to the massless ones). Apart from the general
massive field having two physical degrees of freedom, only so-called
partially massless field of the maximal depth has one degrees of
freedom. This corresponds to the decoupling of the spin-0 component
which happens when
$$
M^2 = m^2 - (s-1)^2 \Lambda = 0
$$

\subsection{Partially massless case of maximal depth}

In this case all $c(k) = 0$ so that all the explicit mass-like terms
vanish and the Lagrangian greatly simplifies:
\begin{eqnarray}
{\cal L}_0 &=& \sum_{k=1}^{s-1} (-1)^{k+1} [ k 
\Omega_{\alpha(2k-1)\beta} e^\beta{}_\gamma 
\Omega^{\alpha(2k-1)\gamma} + \Omega_{\alpha(2k)} D \Phi^{\alpha(2k)}]
\nonumber \\
 && + \sum_{k=1}^{s-2} (-1)^{k+1} b_k [ - \frac{(k+2)}{k} 
\Omega_{\alpha(2)\beta(2k)} e^{\alpha(2)} \Phi^{\beta(2k} + 
\Omega_{\alpha(2k)} e_{\beta(2)} \Phi^{\alpha(2k)\beta(2)} ] \nonumber
\\
 && + E B_{\alpha\beta} B^{\alpha\beta} - B_{\alpha\beta}
e^{\alpha\beta} D A + 2b_0 \Omega_{\alpha(2)} e^{\alpha(2)} A - b_0
f_{\alpha\beta} E^\beta{}_\gamma B^{\alpha\gamma}
\end{eqnarray}
where now:
$$
b_k{}^2 = \frac{k(k+1)(s+k+1)(s-k-1)}{2(k+2)(2k+3)} \Lambda
$$
$$
b_0{}^2 = \frac{(s+1)(s-1)}{3} \Lambda
$$
The Lagrangian is still invariant under all the gauge transformations:
\begin{eqnarray}
\delta_0 \Omega^{\alpha(2k)} &=& D \eta^{\alpha(2k)} + 
\frac{(k+2)b_k}{k} e_{\beta(2)} \eta^{\alpha(2k)\beta(2)} +
\frac{b_{k-1}}{k(2k-1)} e^{\alpha(2)} \eta^{\alpha(2k-2)} \nonumber \\
\delta_0 \Phi^{\alpha(2k)} &=& D \xi^{\alpha(2k)} + e^\alpha{}_\beta
\eta^{\alpha(2k-1)\beta} + b_k e_{\beta(2)} 
\xi^{\alpha(2k)\beta(2)} + \frac{(k+1)b_{k-1}}{k(k-1)(2k-1)}
e^{\alpha(2)} \xi^{\alpha(2k-2)} \nonumber \\
\delta \Omega^{\alpha(2)} &=& D \eta^{\alpha(2)} + 3b_1 e_{\beta(2)}
\eta^{\alpha(2)\beta(2)} \\
\delta \Phi^{\alpha(2)} &=& D \xi^{\alpha(2)} + e^\alpha{}_\beta
\eta^{\alpha\beta} + b_1 e_{\beta(2)} \xi^{\alpha(2)\beta(2)} + 2b_0
e^{\alpha(2)} \xi \nonumber \\
\delta B^{\alpha\beta} &=& 2b_0 \eta^{\alpha\beta}, \qquad
\delta A = D \xi + \frac{b_0}{4} e_{\alpha\beta} \xi^{\alpha\beta}
\nonumber
\end{eqnarray}

Now having in our disposal the explicit form of all gauge
transformations it is not hard to construct a set of gauge invariant
objects for all the fields (two-forms for the one-form fields and
one-forms for the zero-form ones):
\begin{eqnarray}
{\cal R}^{\alpha(2k)} &=& D \Omega^{\alpha(2k)} + \frac{(k+2)b_k}{k}
e_{\beta(2)} \Omega^{\alpha(2k)\beta(2)} + \frac{b_{k-1}}{k(2k-1)}
e^{\alpha(2)} \Omega^{\alpha(2k-2)} \nonumber \\
 {\cal T}^{\alpha(2k)} &=& D \Phi^{\alpha(2k)} + e^\alpha{}_\beta
\Omega^{\alpha(2k-1)\beta} + b_k e_{\beta(2)} 
\Phi^{\alpha(2k)\beta(2)} + \frac{(k+1)b_{k-1}}{k(k-1)(2k-1)}
e^{\alpha(2)} \Phi^{\alpha(2k-2)} \nonumber \\
{\cal R}^{\alpha(2)} &=& D \Omega^{\alpha(2)} + 3b_1 e_{\beta(2)}
\Omega^{\alpha(2)\beta(2)} - \frac{b_0}{2} E^\alpha{}_\beta 
B^{\alpha\beta} \nonumber \\
{\cal T}^{\alpha(2)} &=& D \Phi^{\alpha(2)} + e^\alpha{}_\beta
\Omega^{\alpha\beta} + b_1 e_{\beta(2)} \Phi^{\alpha(2)\beta(2)} +
2b_0 e^{\alpha(2)} A \\
{\cal A} &=& D A - E_{\alpha(2)} B^{\alpha(2)} + \frac{b_0}{4}
e_{\alpha(2)} \Phi^{\alpha(2)} \nonumber \\
{\cal B}^{\alpha(2)} &=& D B^{\alpha(2)} - 2b_0 \Omega^{\alpha(2)} +
3b_1 e_{\beta(2)} B^{\alpha(2)\beta(2)} \nonumber
\end{eqnarray}
As it can be seen from the last line, to achieve gauge invariance we
introduced an extra zero-form $B^{\alpha(4)}$ playing the role of the
Stueckelberg field for the $\eta^{\alpha(4)}$ transformations:
$$
\delta B^{\alpha(4)} = 2b_0 \eta^{\alpha(4)}
$$
But then we have to construct a gauge invariant object for this new
field and this in turn requires introduction of the next extra field
and so on until we exhaust all the gauge symmetries. This results in
the following collection of extra zero-forms $B^{\alpha(2k)}$, $2 \le
k \le s-1$ 
\begin{equation}
\delta B^{\alpha(2k)} = 2b_0 \eta^{\alpha(2k)}
\end{equation}
with the corresponding set of gauge invariant one-forms:
\begin{equation}
{\cal B}^{\alpha(2k)} = D B^{\alpha(2k)} - 2b_0 \Omega^{\alpha(2k)} +
\frac{(k+2)b_k}{k} e_{\beta(2)} B^{\alpha(2k)\beta(2)} + 
\frac{b_{k-1}}{k(2k-1)} e^{\alpha(2)} B^{\alpha(2k-2)}
\end{equation}

In what follows we will need the differential identities for the
two-forms which follows directly from their explicit expressions:
\begin{eqnarray}
D {\cal T}^{\alpha(2k)} &=& - e^\alpha{}_\beta 
{\cal R}^{\alpha(2k-1)\beta} - b_1 e_{\beta(2)} 
{\cal T}^{\alpha(2k)\beta(2)} - \frac{(k+1)b_{k-1}}{k(k-1)(2k-1)}
e^{\alpha(2)} {\cal T}^{\alpha(2k-2)} \nonumber \\
D {\cal T}^{\alpha(2)} &=& - e^\alpha{}_\beta {\cal R}^{\alpha\beta} -
b_1 e_{\beta(2)} {\cal T}^{\alpha(2)\beta(2)} - 2b_0 e^{\alpha(2)}
{\cal A} 
\end{eqnarray}
as well as the similar identities for the one-forms:
\begin{eqnarray}
D {\cal B}^{\alpha(2k)} &=& - 2b_0 {\cal R}^{\alpha(2k)} - 
\frac{(k+2)b_k}{k} e_{\beta(2)} {\cal B}^{\alpha(2k)\beta(2)} -
\frac{b_{k-1}}{k(2k-1)} e^{\alpha(2)} {\cal B}^{\alpha(2k-2)}
\nonumber \\
D {\cal B}^{\alpha(2)} &=& - 2b_0 {\cal R}^{\alpha(2)} - 3b_1
e_{\beta(2)} {\cal B}^{\alpha(2)\beta(2)} 
\end{eqnarray}

Now we have equal numbers of the two-forms ${\cal R}^{\alpha(2k)}$ and
one-forms ${\cal B}^{\alpha(2k)}$, $1 \le k \le s-1$ and we will try
to rewrite the Lagrangian in terms of these objects. For this let us
consider the following ansatz:
\begin{equation}
{\cal L} = \sum_{k=1}^{s-1} (-1)^{k+1} [ e_k {\cal T}_{\alpha(2k)} 
{\cal B}^{\alpha(2k)} + f_k {\cal B}_{\alpha(2k-1)\beta}
e^\beta{}_\gamma {\cal B}^{\alpha(2k-1)\gamma} ]
\end{equation}
Clearly, by construction this Lagrangian is gauge invariant for any
values of the coefficients $e_k$ and $f_k$. But to reproduce our
initial Lagrangian we have to adjust these coefficients so that all
the extra fields decouple. Let us extract all the terms containing
$B^{\alpha(2k)}$:
\begin{eqnarray*}
(-1)^{k+1} \Delta {\cal L} &=& e_k {\cal T}_{\alpha(2k)} D
B^{\alpha(2k)} -  e_{k+1}b_k {\cal T}_{\alpha(2k)\beta(2)}
e^{\beta(2)} B^{\alpha(2k)}  \\
 && - \frac{(k+1)e_{k-1}b_{k-1}}{(k-1)} {\cal T}_{\alpha(2k-2)}
e_{\beta(2)} B^{\alpha(2k-2)\beta(2)} \\
 && - 2f_k D {\cal B}^{\alpha(2k-1)\beta} e_\beta{}^\gamma
B_{\alpha(2k-1)\gamma} - \frac{4(k+2)f_{k+1}b_k}{(k+1)} 
{\cal B}_{\alpha(2k)\beta(2)} E^{\beta(2)} B^{\alpha(2k)} \\
 && - \frac{4(k+1)f_{k-1}b_{k-1}}{(k-1)} {\cal B}_{\alpha(2k-2)}
E_{\beta(2)} B^{\alpha(2k-2)\beta(2)}
\end{eqnarray*}
Now using the differential identities given above we obtain:
\begin{eqnarray*}
- e_k D {\cal T}^{\alpha(2k)} B_{\alpha(2k)} 
 &=& - 2ke_k {\cal R}_{\alpha(2k-1)\beta} e^\beta{}_\gamma
B^{\alpha(2k-1)\gamma} + e_kb_k {\cal T}_{\alpha(2k)\beta(2)}
e^{\beta(2)} B^{\alpha(2k)} \\
 && + \frac{(k+1)e_kb_{k-1}}{(k-1)}
{\cal T}_{\alpha(2k-2)} e_{\beta(2)} B^{\alpha(2k-2)\beta(2)}
\end{eqnarray*}
\begin{eqnarray*}
- 2f_k D {\cal B}^{\alpha(2k-1)\beta} e_\beta{}^\gamma
B_{\alpha(2k-1)\gamma} &=& 4b_0f_k {\cal R}^{\alpha(2k-1)\beta} 
e_\beta{}^\gamma B_{\alpha(2k-1)\gamma} \\
 && + \frac{4(k+2)f_kb_k}{k} {\cal B}^{\alpha(2k)\beta(2)}
E_{\beta(2)} B_{\alpha(2k)} \\
 && + \frac{4(k+1)f_kb_{k-1}}{k} {\cal B}^{\alpha(2k-2)} E^{\beta(2)}
B_{\alpha(2k-2)\beta(2)}
\end{eqnarray*}
Collecting all pieces together we finally get:
\begin{eqnarray}
(-1)^{k+1} \Delta {\cal L} &=& (4b_0f_k - 2ke_k) 
{\cal R}_{\alpha(2k-1)\beta} e^\beta{}_\gamma B^{\alpha(2k-1)\gamma}
\nonumber \\
 && + (e_k-e_{k+1})b_k {\cal T}_{\alpha(2k)\beta(2)} e^{\beta(2)}
B^{\alpha(2k)} \nonumber \\
 && + \frac{(k+1)b_{k-1}}{(k+1)}(e_k-e_{k-1}) {\cal T}_{\alpha(2k-2)}
e_{\beta(2)} B^{\alpha(2k-2)\beta(2)} \nonumber \\
 && + 4(k+2)b_k(\frac{f_k}{k}-\frac{f_{k+1}}{(k+1)}) 
{\cal B}_{\alpha(2k)\beta_2)} e^{\beta(2)} B^{\alpha(2k)} \nonumber \\
 && + 4(k+1)b_{k-1} (\frac{f_k}{k} - \frac{f_{k-1}}{(k-1)})
{\cal B}_{\alpha(2k-2)} e_{\beta(2)} B^{\alpha(2k-2)\beta(2)}
\end{eqnarray}
Note that to have a correct normalization of the kinetic terms for all
fields we have to put
\begin{equation}
e_k = - \frac{1}{2b_0}, \qquad
f_k = - \frac{k}{4b_0{}^2}
\end{equation}
and it is easy to see the in this case all the terms containing
$B^{\alpha(2k)}$ above vanish. 

To complete, we have to consider terms with the $B^{\alpha(2)}$ field.
We get:
\begin{eqnarray*}
\Delta {\cal L} &=&
e_1 {\cal T}_{\alpha(2)} D B^{\alpha(2)} - b_1e_2
{\cal T}_{\alpha(2)\beta(2)} e^{\beta(2)} B^{\alpha(2)} \\
 && + 2f_1 D {\cal B}^{\alpha\beta} e_\beta{}^\gamma B_{\alpha\gamma}
- 6b_1f_2 {\cal B}_{\alpha(2)\beta(2)} E^{\beta(2)} B^{\alpha(2)}
\end{eqnarray*}
Once again using the differential identities:
\begin{eqnarray*}
- e_1 D {\cal T}^{\alpha(2)} B_{\alpha(2)} &=& e_1 [ e^\alpha{}_\beta
{\cal R}^{\alpha\beta} + b_1 e_{\beta(2)} {\cal T}^{\alpha(2)\beta(2)}
+ 2b_0  e^{\alpha(2)} {\cal A} ] B_{\alpha(2)} \\
 2f_1 D {\cal B}^{\alpha\beta} e_\beta{}^\gamma B_{\alpha\gamma}
&=& - 4f_1b_0 {\cal R}^{\alpha\beta} e_\beta{}^\gamma B_{\alpha\gamma}
+ 12b_1f_1 {\cal B}^{\alpha(2)\beta(2)} E_{\beta(2)} B_{\alpha(2)}
\end{eqnarray*}
we finally obtain:
\begin{eqnarray}
\Delta {\cal L} &=& (2e_1 - 4f_1b_0) {\cal R}^{\alpha\beta} 
e^\beta{}_\gamma B_{\alpha\gamma} + b_1(e_1-e_2) 
{\cal T}_{\alpha(2)\beta(2)} e^{\beta(2)} B^{\alpha(2)} \nonumber \\
 && + 6b_1(2f_1-f_2) {\cal B}_{\alpha(2)\beta(2)} e^{\beta(2)}
B^{\alpha(2)} + 2b_0e_1 e^{\alpha(2)} B_{\alpha(2)} {\cal A} \nonumber
\\
 &=& 2b_0e_1 e^{\alpha(2)} B_{\alpha(2)} {\cal A}
\end{eqnarray}

Thus for the chosen coefficients $e_k$ and $f_k$ we achieved complete
decoupling for all the extra fields and the resulting Lagrangian (as we
have explicitly checked) reproduces our initial one. Note at last,
that this Lagrangian can be rewritten also in the more common to three
dimensions CS-like form:
\begin{equation}
{\cal L} = \sum_{k=1}^{s-1} (-1)^{k+1} [ {\cal T}_{\alpha(2k)}
\Omega^{\alpha(2k)} - k \Omega_{\alpha(2k-1)\beta} e^\beta{}_\gamma
\Omega^{\alpha(2k-1)\gamma} ] - e^{\alpha(2)} B_{\alpha(2)} {\cal A}
\end{equation}

\subsection{Partial gauge fixing}

In the frame-like formalism for the massless spin-$s$ field 
($\Omega^{\alpha(2s-2)}$, $\Phi^{\alpha(2s-2)}$) in anti-de Sitter
space ($\Lambda = - \lambda^2 < 0$) there exists a very convenient
possibility of the separation of variables. Namely, let us consider
the Lagrangian for such massless field:
\begin{eqnarray}
{\cal L}_0 &=& (-1)^s [ (s-1) \Omega_{\alpha(2s-3)\beta} 
e^\beta{}_\gamma \Omega^{\alpha(2s-3)\gamma} + \Omega_{\alpha(2s-2)} D
\Phi^{\alpha(2s-2)} \nonumber \\
 && \qquad + \frac{(s-1)\lambda^2}{4} \Phi_{\alpha(2s-3)\beta}
e^\beta{}_\gamma \Phi^{\alpha(2s-3)\gamma} ]
\end{eqnarray}
This Lagrangian is invariant under the following local gauge
transformations:
\begin{eqnarray}
\delta \Omega^{\alpha(2s-2)} &=& D \eta^{\alpha(2s-2)} +
\frac{\lambda^2}{4} e^\alpha{}_\beta \xi^{\alpha(2s-3)\beta} \nonumber
\\
\delta \Phi^{\alpha(2s-2)} &=& D \xi^{\alpha(2s-2)} + e^\alpha{}_\beta
\eta^{\alpha(2s-3)\beta}
\end{eqnarray}

Let us introduce new variables:
\begin{eqnarray}
\hat{\Omega}^{\alpha(2s-2)} &=& \Omega^{\alpha(2s-2)} + 
\frac{\lambda}{2} \Phi^{\alpha(2s-2)} \nonumber \\
\hat{\Phi}^{\alpha(2s-2)} &=& \Omega^{\alpha(2s-2)} - 
\frac{\lambda}{2} \Phi^{\alpha(2s-2)}
\end{eqnarray}
and similarly for the parameters of the gauge transformations:
\begin{eqnarray}
\hat{\eta}^{\alpha(2s-2)} &=& \eta^{\alpha(2s-2)} + \frac{\lambda}{2}
\xi^{\alpha(2s-2)} \nonumber \\
\hat{\xi}^{\alpha(2s-2)} &=& \eta^{\alpha(2s-2)} - \frac{\lambda}{2}
\xi^{\alpha(2s-2)}
\end{eqnarray}
Then the Lagrangian can be rewritten as the sum of the two
independent parts:
\begin{eqnarray}
{\cal L}_0 &=& \frac{(-1)^s}{2\lambda} [ (s-1)\lambda 
\hat{\Omega}_{\alpha(2s-3)\beta} e^\beta{}_\gamma 
\hat{\Omega}^{\alpha(2s-3)\gamma} + \hat{\Omega}_{\alpha(2s-2)} D
\hat{\Omega}^{\alpha(2s-2)} \nonumber \\
 && \qquad + (s-1)\lambda \hat{\Phi}_{\alpha(2s-3)\beta} 
e^\beta{}_\gamma \hat{\Phi}^{\alpha(2s-3)\gamma} - 
\hat{\Phi}_{\alpha(2s-2)} D \hat{\Phi}^{\alpha(s-2)} ] 
\end{eqnarray}
while the gauge transformations take the form:
\begin{eqnarray}
\delta \hat{\Omega}^{\alpha(2s-2)} &=& D \hat{\eta}^{\alpha(2s-2)} + 
\frac{\lambda}{2} e^\alpha{}_\beta \hat{\eta}^{\alpha(2s-3)\beta}
\nonumber \\
\delta \hat{\Phi}^{\alpha(2s-2)} &=& D \hat{\xi}^{\alpha(2s-2)} -
\frac{\lambda}{2} e^\alpha{}_\beta \hat{\xi}^{\alpha(2s-3)\beta}
\end{eqnarray}
Moreover, such separation of variables works not only in the free case
but in the interacting case as well.

As we have shown previously \cite{BSZ12a}, similar mechanism is
possible for the massive fields provided one uses a partial gauge
fixing. Namely, let us return to the general massive case, described
above, and use the $\xi$ gauge transformation to set the gauge
$\varphi = 0$. Then, solving the spin-0 equation
$$
A = \frac{1}{2Ms} e_{\alpha(2)} \pi^{\alpha(2)}
$$
we obtain the following Lagrangian (after rescaling $\pi \Rightarrow
2Ms\pi$):
\begin{eqnarray}
{\cal L} &=& \sum_{k=1}^{s-1} (-1)^{k+1} [ k 
\Omega_{\alpha(2k-1)\beta} e^\beta{}_\gamma 
\Omega^{\alpha(2k-1)\gamma} + \Omega_{\alpha(2k)} D \Phi^{\alpha(2k)}
+ c_k \Phi_{\alpha(2k-1)\beta} e^\beta{}_\gamma 
\Phi^{\alpha(2k-1)\gamma} ] \nonumber \\
 && + \sum_{k=1}^{s-2} (-1)^{k+1} b_k [ - \frac{(k+2)}{k} 
\Omega_{\alpha(2)\beta(2k)} e^{\alpha(2)} \Phi^{\beta(2k)} + 
\Omega_{\alpha(2k)} e_{\beta(2)} \Phi^{\alpha(2k)\beta(2)} ] \nonumber
\\
 && + E B_{\alpha\beta} B^{\alpha\beta} + 4  B_{\alpha\gamma} 
E^{\alpha\beta} D \pi_\beta{}^\gamma + 4M^2s^2 E 
\pi_{\alpha\beta} \pi^{\alpha\beta} \nonumber \\
 && + 8b_0 \Omega_{\alpha\gamma} E^{\alpha\beta} \pi_\beta{}^\gamma -
b_0 \Phi_{\alpha\beta} E^\beta{}_\gamma B^{\alpha\gamma}
\end{eqnarray}
Now let us introduce new variables:
\begin{eqnarray}
\hat{\Omega}^{\alpha(2k)} &=& \Omega^{\alpha(2k)} + \frac{Ms}{2k(k+1)}
\Phi^{\alpha(2k)} \nonumber \\
\hat{\Phi}^{\alpha(2k)} &=& \Omega^{\alpha(2k)} - \frac{Ms}{2k(k+1)}
\Phi^{\alpha(2k)}
\end{eqnarray}
and similarly for the zero-forms:
\begin{eqnarray}
\hat{B}^{\alpha(2)} &=& B^{\alpha(2)} + 2Ms \pi^{\alpha(2)} \nonumber
\\
\hat{\pi}^{\alpha(2)} &=& B^{\alpha(2)} - 2Ms \pi^{\alpha(2)} 
\end{eqnarray}
then the Lagrangian also decomposes into two independent parts, one
for the fields ($\hat{\Omega}$, $\hat{B}$) and the other one for the
fields ($\hat{\Phi}$, $\hat{\pi}$). From now on, we consider the first
part only. The Lagrangian has the form (omitting hats):
\begin{eqnarray}
{\cal L} &=& \sum_{k=1}^{s-1} (-1)^{k+1} [ \frac{k}{2}Ms
\Omega_{\alpha(2k-1)\beta} e^\beta{}_\gamma 
\Omega^{\alpha(2k-1)\gamma} + \frac{k(k+1)}{2} \Omega_{\alpha(2k)} D
\Omega^{\alpha(2k)} ] \nonumber \\
 && - \sum_{k=1}^{s-2} (-1)^{k+1} (k+1)(k+2)b_k 
\Omega_{\alpha(2k)\beta(2)} e^{\beta(2)} \Omega^{\alpha(2k)} \nonumber
\\
 && + \frac{Ms}{2} E B_{\alpha(2)} B^{\alpha(2)} + \frac{1}{2}
B_{\alpha\gamma} E^{\alpha\beta} D B_\beta{}^\gamma + 2b_0 
\Omega_{\alpha\gamma} E^{\alpha\beta} B_\beta{}^\gamma
\end{eqnarray}
This Lagrangian is invariant under its own half of the gauge
transformations:
\begin{eqnarray}
\delta \Omega^{\alpha(2k)} &=& D \eta^{\alpha(2k)} +
\frac{Ms}{2k(k+1)} e^\alpha{}_\beta \eta^{\alpha(2k-1)\beta}
\nonumber \\
 && + \frac{(k+2)}{k}b_k e_{\beta(2)} \eta^{\alpha(2k)\beta(2)} 
+ \frac{b_{k-1}}{k(2k-1)} e^{\alpha(2)} \eta^{\alpha(2k-2)}
\nonumber \\
\delta \Omega^{\alpha(2)} &=& D \eta^{\alpha(2)} + 
\frac{Ms}{4} e^\alpha{}_\beta \eta^{\alpha\beta} + 3b_1
e_{\beta(2)} \eta^{\alpha(2)\beta(2)} \\
\delta B^{\alpha(2)} &=& 2b_0 \eta^{\alpha(2)} \nonumber
\end{eqnarray}

Now we proceed with the construction of the gauge invariant objects
for all the fields:
\begin{eqnarray}
{\cal R}^{\alpha(2k)} &=& D \Omega^{\alpha(2k)} + \frac{Ms}{2k(k+1)}
e^\alpha{}_\beta \Omega^{\alpha(2k-1)\beta} + \frac{(k+2)}{k}b_k
e_{\beta(2)} \Omega^{\alpha(2k)\beta(2)} \nonumber \\
 && + \frac{b_{k-1}}{k(2k-1)} e^{\alpha(2)} \Omega^{\alpha(2k-2)}
\nonumber \\
{\cal R}^{\alpha(2)} &=& D \Omega^{\alpha(2)} + \frac{Ms}{4} 
e^\alpha{}_\beta \Omega^{\alpha\beta} + 3b_1 e_{\beta(2)}
\Omega^{\alpha(2)\beta(2)} - \frac{b_0}{2} E^\alpha{}_\beta 
B^{\alpha\beta} \\
{\cal B}^{\alpha(2)} &=& D B^{\alpha(2)} - 2b_0 \Omega^{\alpha(2)} +
\frac{Ms}{4} e^\alpha{}_\beta B^{\alpha\beta} + 3b_1 e_{\beta(2)}
B^{\alpha(2)\beta(2)} \nonumber
\end{eqnarray}
Again, to achieve the gauge invariance for the ${\cal B}^{\alpha(2)}$
we introduced an extra field $B^{\alpha(4)}$:
$$
\delta B^{\alpha(4)} = 2b_0 \eta^{\alpha(4)}
$$
As in the previous case, the procedure ends up with the set of such
extra zero-forms $B^{\alpha(2k)}$, $2 \le k \le s-1$
\begin{equation}
\delta B^{\alpha(2k)} = 2b_0 \eta^{\alpha(2k)}
\end{equation}
with the appropriate set of gauge invariant one-forms:
\begin{eqnarray}
{\cal B}^{\alpha(2k)} &=& D B^{\alpha(2k)} - 2b_0 \Omega^{\alpha(2k)}
+ \frac{Ms}{2k(k+1)} e^\alpha{}_\beta B^{\alpha(2k-1)\beta} \nonumber
\\
 && + \frac{b_{k-1}}{k(2k-1)} e^{\alpha(2)} B^{\alpha(2k-2)} 
 + \frac{(k+2)}{k}b_k e_{\beta(2)} B^{\alpha(2k)\beta(2)}
\end{eqnarray}

In what follows we will need the differential identities for the
two-forms only:
\begin{eqnarray}
D {\cal R}^{\alpha(2k)} &=& - \frac{Ms}{2k(k+1)} e^\alpha{}_\beta
{\cal R}^{\alpha(2k-1)\beta} - \frac{(k+2)}{k}b_k e_{\beta(2)}
{\cal R}^{\alpha(2k)\beta(2)} \nonumber \\
 && - \frac{b_{k-1}}{k(2k-1)} e^{\alpha(2)}
{\cal R}^{\alpha(2k-2)} \\
D {\cal R}^{\alpha(2)} &=& - \frac{Ms}{4} e^\alpha{}_\beta 
{\cal R}^{\alpha\beta} - 3b_1 e_{\beta(2)} 
{\cal R}^{\alpha(2)\beta(2)} - \frac{b_0}{2} E^\alpha{}_\beta
{\cal B}^{\alpha\beta} \nonumber
\end{eqnarray}

Now let us try to rewrite the Lagrangian in the explicitly gauge
invariant form. It happens that in this case it is enough to consider
the following ansatz:
\begin{equation}
{\cal L} = \sum (-1)^k e_k {\cal R}_{\alpha(2k)} {\cal B}^{\alpha(2k)}
\end{equation}
Indeed, let us extract all the terms with the extra field
$B^{\alpha(2k)}$:
\begin{eqnarray*}
(01)^{k+1} \Delta {\cal L} &=& e_k [ {\cal R}_{\alpha(2k)} D
B^{\alpha(2k)} + \frac{Ms}{(k+1)} {\cal R}_{\alpha(2k-1)\beta} 
e^\beta{}_\gamma B^{\alpha(2k-1)\gamma} ] \\
 && - e_{k+1}b_k {\cal R}_{\alpha(2k)\beta(2)} e^{\beta(2)}
B^{\alpha(2k)} \\
 && - \frac{(k+1)}{(k-1)}b_{k-1}e_{k-1} {\cal R}_{\alpha(2k-2)} 
e_{\beta(2)} B^{\alpha(2k-2)\beta(2)}
\end{eqnarray*}
One more time using the differential identity:
\begin{eqnarray*}
- e_k D {\cal R}^{\alpha(2k)} B_{\alpha(2k)} &=& -
\frac{Ms}{(k+1)}e_k {\cal R}_{\alpha(2k-1)\beta} e^\beta{}_\gamma
B^{\alpha(2k-1)\gamma} \\
 && + \frac{(k+2)}{k}e_kb_k   {\cal R}_{\alpha(2k)\beta(2)}
e^{\beta(2)} B^{\alpha(2k)} \\
 && + e_kb_{k-1} {\cal R}_{\alpha(2k-2)} e_{\beta(2)} 
B^{\alpha(2k-2)\beta(2)}
\end{eqnarray*}
we obtain finally:
\begin{eqnarray}
(-1)^{k+1} \Delta {\cal L} &=& (\frac{(k+2)}{k}e_k-e_{k+1})b_k 
{\cal R}_{\alpha(2k)\beta(2)} e^{\beta(2)} B^{\alpha(2k)} \nonumber \\
 && + (e_k-\frac{(k+1)}{(k-1)}e_{k-1})b_{k-1} {\cal R}_{\alpha(2k-2)}
e_{\beta(2)} B^{\alpha(2k-2)\beta(2)}
\end{eqnarray}
In this case to have the correct normalization for the kinetic terms
we have to put:
\begin{equation}
e_k = - \frac{k(k+1)}{4b_0} \qquad \Rightarrow \qquad
e_{k-1} = \frac{(k-1)}{(k+1)} e_k
\end{equation}
and it is easy to see that all the terms with the $B^{\alpha(2k)}$
fields vanish.

To complete, we have to consider terms with the $B^{\alpha(20}$ field
as well. We get:
$$
\Delta {\cal L} =
- e_1 D {\cal R}_{\alpha(2)} B^{\alpha(2)} + \frac{Mse_1}{2}
{\cal R}_{\alpha\beta} e^\beta{}_\gamma B^{\alpha\gamma} - b_1e_2
{\cal R}_{\alpha(2)\beta(2)} e^{\beta(2)} B^{\alpha(2)}
$$
while the differential identity leads to
$$
- e_1 D {\cal R}^{\alpha(2)} B_{\alpha(2)} = - \frac{Mse_1}{4}
{\cal R}_{\alpha\beta} e^\beta{}_\gamma B_{\alpha\gamma} + 3b_1e_1
{\cal R}_{\alpha(2)\beta(2)} e^{\beta(2)} B^{\alpha(2)} - 
b_0e_1 E^\alpha{}_\beta {\cal B}^{\beta\gamma} B_{\alpha\gamma}
$$
Thus we obtain:
\begin{eqnarray}
\Delta {\cal L} &=& (3e_1-e_2)b_1 {\cal R}_{\alpha(2)\beta(2)}
e^{\beta(2)} B^{\alpha(2)} - b_0e_1 E^\alpha{}_\beta
{\cal B}^{\beta\gamma} B_{\alpha\gamma} \nonumber \\
 &=& - b_0e_1 E^\alpha{}_\beta {\cal B}^{\beta\gamma} B_{\alpha\gamma}
\end{eqnarray}

So our ansatz with the coefficients $e_k$ given above does reproduce
our initial Lagrangian (as we have explicitly checked). Due to the
decoupling of all the extra fields this Lagrangian can be also
rewritten in the CS-like form:
\begin{equation}
{\cal L} = \sum_{k=1}^{s-1} (-1)^{k+1} \frac{k(k+1)}{2}
{\cal R}_{\alpha(2k)} \Omega^{\alpha(2k)} + \frac{1}{2}
E^\alpha{}_\beta {\cal B}^{\beta\gamma} B_{\alpha\gamma}
\end{equation}

\section{Half-integer spin}

In this section we consider a massive fermionic field with
half-integer spin $s+1/2$. The frame-like gauge invariant formulation
\cite{BSZ14a} requires a set of one-forms $\Phi^{\alpha(2k+1)}$, $0
\le k \le s-1$, as well as zero-form $\phi^\alpha$. The Lagrangian
describing such massive field in $(A)dS_3$ background has the form:
\begin{eqnarray}
\frac{1}{i} {\cal L} &=& \sum_{k=0}^{s-1} (-1)^{k+1} [ \frac{1}{2} 
\Phi_{\alpha(2k+1)} D \Phi^{\alpha(2k+1)} ]
+ \frac{1}{2} \phi_\alpha E^\alpha{}_\beta D \phi^\beta \nonumber \\
 && + \sum_{k=1}^{s-1} (-1)^{k+1} a_k \Phi_{\alpha(2k-1)\beta(2)}
e^{\beta(2)} \Phi^{\alpha(2k-1)} + a_0 \Phi_\alpha E^\alpha{}_\beta
\phi^\beta \nonumber \\
 && + \sum_{k=0}^{s-1} (-1)^{k+1} \frac{b_k}{2}
\Phi_{\alpha(2k)\beta} e^\beta{}_\gamma \Phi^{\alpha(2k)\gamma}
- \frac{3b_0}{2} E \phi_\alpha \phi^\alpha
\end{eqnarray}
where
$$
b_k = \frac{(2s+1)}{(2k+3)} M, \qquad
M^2 = m^2 - (s-\frac{1}{2})^2 \Lambda
$$
$$
a_k{}^2 = \frac{(s+k+1)(s-k)}{2(k+1)(2k+1)} [ M^2 + (2k+1)^2
\frac{\Lambda}{4} ]
$$
$$
a_0{}^2 = 2s(s+1) [ M^2 + \frac{\Lambda}{4} ]
$$
The structure of this Lagrangian repeats the general pattern: the
first and the third lines are the sum of the kinetic and mass-like
terms for all the components, while the second line contains all
necessary cross-terms.

This Lagrangian is invariant under the following gauge transformations
with the fermionic parameters:
\begin{eqnarray}
\delta \Phi^{\alpha(2k+1)} &=& D \xi^{\alpha(2k+1)} + 
\frac{b_k}{(2k+1)} e^\alpha{}_\beta \xi^{\alpha(2k)\beta} \nonumber \\
 && + \frac{a_k}{k(2k+1)} e^{\alpha(2)} \xi^{\alpha(2k-1)} 
 + a_{k+1} e_{\beta(2)} \xi^{\alpha(2k+1)\beta(2)} \\
\delta \phi^\alpha &=& a_0 \xi^\alpha \nonumber
\end{eqnarray}

Now we construct a set of gauge invariant objects for all fields
(two-forms for $\Phi$ and one-form for $\phi$):
\begin{eqnarray}
{\cal R}^{\alpha(2k+1)} &=& D \Phi^{\alpha(2k+1)} + \frac{b_k}{(2k+1)}
e^\alpha{}_\beta \Phi^{\alpha(2k)\beta} + \frac{a_k}{k(2k+1)}
e^{\alpha(2)} \Phi^{\alpha(2k-1)} \nonumber \\
 && + a_{k+1} e_{\beta(2)} \Phi^{\alpha(2k+1)\beta(2)} \nonumber \\
{\cal R}^\alpha &=& D \Phi^\alpha + b_0 e^\alpha{}_\beta \Phi^\beta +
a_1 e_{\beta(2)} \Phi^{\alpha\beta(2)} - a_0 E^\alpha{}_\beta
\phi^\beta \\
{\cal F}^\alpha &=& D \phi^\alpha - a_0 \Phi^\alpha + b_0 
e^\alpha{}_\beta \phi^\beta + a_1 e_{\beta(2)} 
\phi^{\alpha\beta(2)} \nonumber
\end{eqnarray}
Similarly to the bosonic case, to achieve gauge invariance for the
last one we introduced an extra zero-form $\phi^{\alpha(3)}$ playing
the role of the Stueckelberg field for the $\xi^{\alpha(3)}$
transformations:
$$
\delta \phi^{\alpha(3)} = a_0 \xi^{\alpha(3)}
$$
The whole procedure ends with the set of such extra zero-forms
$\phi^{\alpha(2k+1)}$, $1 \le k \le s-1$ with the appropriate gauge
invariant one-forms:
\begin{eqnarray}
{\cal F}^{\alpha(2k+1)} &=& D \phi^{\alpha(2k+1)} - a_0 
\Phi^{\alpha(2k+1)} + \frac{b_k}{(2k+1)} e^\alpha{}_\beta
\phi^{\alpha(2k)\beta} \nonumber \\
 && + \frac{a_k}{k(2k+1)} e^{\alpha(2)} \phi^{\alpha(2k-1)} + a_{k+1}
e_{\beta(2)} \phi^{\alpha(2k+1)\beta(2)}
\end{eqnarray}
where
\begin{equation}
\delta \phi^{\alpha(2k+1)} = a_0 \xi^{\alpha(2k+1)}
\end{equation}

In what follows we will need the differential identities for the 
two-forms:
\begin{eqnarray}
D {\cal R}^{\alpha(2k+1)} &=& - \frac{b_k}{(2k+1)} e^\alpha{}_\beta
{\cal R}^{\alpha(2k)\beta} - \frac{a_k}{k(2k+1)} e^{\alpha(2)}
{\cal R}^{\alpha(2k-1)} \nonumber \\
 && - a_{k+1} e_{\beta(2)} {\cal R}^{\alpha(2k+1)\beta(2)} \\
D {\cal R}^\alpha &=& - b_0 e^\alpha{}_\beta {\cal R}^\beta - a_1
e_{\beta(2)} {\cal R}^{\alpha\beta(2)} - a_0 E^\alpha{}_\beta
{\cal F}^\beta \nonumber
\end{eqnarray}

Now having in our disposal an equal number of two-forms and one-forms
we will try to rewrite the Lagrangian in the explicitly gauge
invariant form. For this we consider the following ansatz:
\begin{equation}
{\cal L} = \sum_{k=0}^{s-1} (-1)^{k+1} e_k {\cal R}_{\alpha(2k+1)}
{\cal T}^{\alpha(2k+1)}
\end{equation}
We have to adjust the coefficients $e_k$ in such a way so that all
extra zero-forms decouple. Let us extract all the terms containing the
zero-form $\phi^{\alpha(2k+1)}$:
\begin{eqnarray*}
(-1)^{k+1} \Delta {\cal L} &=& e_k D {\cal R}^{\alpha(2k+1)}
\phi_{\alpha(2k+1)} + e_kb_k {\cal R}_{\alpha(2k)\beta} 
e^\beta{}_\gamma \phi^{\alpha(2k)\gamma} \\
 && - e_{k+1}a_{k+1} {\cal R}_{\alpha(2k+1)\beta(2)} e^{\beta(2)}
\phi^{\alpha(2k+1)} \\
 && - e_{k-1}a_k {\cal R}_{\alpha(2k-1)} e_{\beta(2)}
\phi^{\alpha(2k-1)\beta(2)}
\end{eqnarray*}
With the help of the differential identity for 
${\cal R}^{\alpha(2k+1)}$ that gives:
\begin{eqnarray*}
e_k D {\cal R}^{\alpha(2k+1)} \phi_{\alpha(2k+1)} 
 &=& - e_kb_k {\cal R}_{\alpha(2k)\beta} e^\beta{}_\gamma
\phi^{\alpha(2k)\gamma} \\
 && + e_ka_{k+1} {\cal R}_{\alpha(2k+1)\beta(2)}
e^{\beta(2)} \phi^{\alpha(2k+1)} \\
 && + e_ka_k {\cal R}_{\alpha(2k-1)}  e_{\beta(2)} 
\phi^{\alpha(2k-1)\beta(2)}
\end{eqnarray*}
we obtain finally:
\begin{eqnarray}
\Delta {\cal L} &=& (e_k-e_{k+1})a_{k+1} 
{\cal R}_{\alpha(2k+1)\beta(2)} e^{\beta(2)} \phi^{\alpha(2k+1)}
\nonumber \\
 && + (e_k-e_{k-1})a_k  {\cal R}_{\alpha(2k-1)} e_{\beta(2)}
\phi^{\alpha(2k-1)\beta(2)}
\end{eqnarray}
In this case to get the correct normalization for the kinetic terms we
have to put:
\begin{equation}
e_k = - \frac{i}{2a_0}
\end{equation}
and as a result all the terms containing extra fields vanish.

To complete we have to consider the terms with $\phi^\alpha$:
$$
\Delta {\cal L} = e_0 D {\cal R}^\alpha \phi_\alpha + e_0b_0 
{\cal R}_\alpha e^\alpha{}_\beta \phi^\beta - e_1a_1 
{\cal R}_{\alpha\beta(2)} e^{\beta(2)} \phi^\alpha
$$
Once again using the differential identity
$$
e_0 D {\cal R}^\alpha \phi_\alpha = - e_0b_0 {\cal R}_\alpha
e^\alpha{}_\beta \phi^\beta + e_0a_1 {\cal R}_{\alpha\beta(2)}
e^{\beta(2)} \phi^\alpha + e_0a_0 {\cal F}_\alpha E^\alpha{}_\beta
\phi^\beta
$$
we obtain:
\begin{eqnarray}
\Delta {\cal L} &=& (e_0-e_1)a_1 {\cal R}_{\alpha\beta(2)}
e^{\beta(2)} \phi^\alpha + e_0a_0 {\cal F}_\alpha E^\alpha{}_\beta
\phi^\beta \nonumber \\
 &=& e_0a_0 {\cal F}_\alpha E^\alpha{}_\beta \phi^\beta 
\end{eqnarray}
Thus we obtained the explicitly gauge invariant expression that
correctly reproduce our Lagrangian (as we have checked). As the last
remark note that this Lagrangian also can be rewritten in the CS-like
form:
\begin{equation}
\frac{1}{i} {\cal L} = \sum_{k=0}^{s-1} (-1)^{k+1} \frac{1}{2} 
{\cal R}_{\alpha(2k+1)} \Phi^{\alpha(2k+1)} - \frac{1}{2} 
{\cal F}_\alpha E^\alpha{}_\beta \phi^\beta
\end{equation}

\section*{Conclusion}

Thus we have shown that using the frame-like gauge invariant formalism
for the massive (and partially massless) higher spin fields in three
dimensions the free Lagrangian for these fields can be rewritten in
the explicitly gauge invariant form it terms of the appropriately
chosen set of gauge invariant objects. These gauge invariant objects
necessarily contains not only Lagrangian fields but a set of extra
fields as well and these extra fields play an important role in the
interacting theories. At the same time the free Lagrangians are
constructed in such a way that all these extra fields decouple.

An open question is how to deal with the systems containing both
massless and massive higher spins because the massless fields are
naturally described bu the Chern-Simons theories. Note however that in
both cases (massless and massive) we have a set of gauge invariant
objects. So we certainly can proceed with the first stage of
FV-formalism, i.e. with the deformation procedure. But the
construction of the Lagrangian formalism for the interacting theories
is still has to be developed.

\section*{Acknowledgments}

Author is grateful to I.~L.~Buchbinder and T.~V.~Snegirev for
collaboration. The author acknowledge a kind hospitality extended to
him at the MIAPP program "Higher Spin Theory and Duality" (Munich,
May 2-27 2016) where this talk was given. Work was supported in parts
by RFBR grant No. 14-02-01172.

\appendix

\section{Massive spin-2 in $d \ge 3$}

In this appendix, using the massive spin-2 case as an example, we
illustrate the relation of our three-dimensional results with their
higher-dimensiohnal analougs.

The frame-like gauge invariant formulation for the massive spin-2
field \cite{Zin08b} requires three pairs of auxiliary and physical
fields: ($\Omega_\mu{}^{ab}$, $f_\mu{}^a$), ($B^{ab}$, $B_\mu$) and 
($\pi^a$, $\varphi$). The Lagrangian describing such massive field in
$(A)dS_d$ background ($d \ge 3$) has the form:
\begin{eqnarray} 
{\cal L}_0 &=& \frac{1}{2} \eptwo \Omega_\mu{}^{ac} \Omega_\nu{}^{bc}
- \frac{1}{2} \epthree \Omega_\mu{}^{ab} D_\nu f_\alpha{}^c + 
\frac{1}{2} B_{ab}{}^2 \nonumber \\
 && - \eptwo B^{ab} D_\mu B_\nu - \frac{(d-2)}{2(d-1)} \pi_a{}^2 +
\frac{(d-2)}{(d-1)} e^\mu{}_a \pi^a D_\mu \varphi \nonumber \\
&& + m [ \eptwo \Omega_\mu{}^{ab} B_\nu + e^\mu{}_a B^{ab} f_\mu{}^b
] - 2M e^\mu{}_a \pi^a B_\mu \nonumber \\
&& + \frac{M^2}{2} \eptwo f_\mu{}^a f_\nu{}^b - mM e^\mu{}_a f_\mu{}^a
\varphi + \frac{d}{2(d-1)} m^2 \varphi^2
\end{eqnarray}
where
\begin{equation}
M^2 = m^2 - \kappa(d-2)
\end{equation}
This Lagrangian is invariant under the following gauge
transformations:
\begin{eqnarray}
\delta \Omega_\mu{}^{ab} &=& D_\mu \eta^{ab}
- \frac{M^2}{(d-2)} e_\mu{}^{[a} \xi^{b]} \nonumber \\
\delta f_\mu{}^a &=& D_\mu \xi^a + \eta_\mu{}^a + \frac{2m}{(d-2)}
e_\mu{}^a \xi \nonumber  \\
\delta B_\mu &=& D_\mu \xi + \frac{m}{2} \xi_\mu, \qquad
\delta B^{ab} = - m \eta^{ab} \\
\delta \varphi &=& \frac{2(d-1)}{(d-2)} M \xi, \qquad
\delta \pi^a = - \frac{(d-1)}{(d-2)} mM \xi^a \nonumber
\end{eqnarray}

For all the six fields we can construct the gauge invariant objects
(two-forms or one-forms):
\begin{eqnarray}
{\cal F}_{\mu\nu}{}^{ab} &=& D_{[\mu} \Omega_{\nu]}{}^{ab} -
\frac{m}{(d-2)} e_{[\mu}{}^{[a} B_{\nu]}{}^{b]} - \frac{M^2}{(d-2)} 
e_{[\mu}{}^{[a} f_{\nu]}{}^{b]} + \frac{2mM}{(d-1)(d-2)}
e_{[\mu}{}^a e_{\nu]}{}^b \varphi \nonumber \\
T_{\mu\nu}{}^a &=& D_{[\mu} f_{\nu]}{}^a - \Omega_{[\mu,\nu]}{}^a +
\frac{2m}{(d-2)} e_{[\mu}{}^a B_{\nu]} \nonumber \\
{\cal B}_\mu{}^{ab} &=& D_\mu B^{ab} + m \Omega_\mu{}^{ab} -
\frac{M}{(d-1)} e_\mu{}^{[a} \pi^{b]} \nonumber \\
{\cal B}_{\mu\nu} &=& D_{[\mu} B_{\nu]} - B_{\mu\nu} - \frac{m}{2}
f_{[\mu,\nu]} \\
\Pi_\mu{}^a &=& D_\mu \pi^a + \frac{(d-1)}{(d-2)} M B_\mu{}^a
+ \frac{(d-1)}{(d-2)} mM f_\mu{}^a - \frac{m^2}{(d-2)} e_\mu{}^a
\varphi \nonumber \\
\Phi_\mu &=& D_\mu \varphi - \pi_\mu - \frac{2(d-1)}{(d-2)} M B_\mu
\nonumber
\end{eqnarray}
They satisfy the following differential identities:
\begin{eqnarray}
D_{[\mu} {\cal F}_{\nu\alpha]}{}^{ab} &=& \frac{m}{(d-2)} 
e_{[\mu}{}^{[a} {\cal B}_{\nu,\alpha]}{}^{b]} + \frac{M^2}{(d-2)}
e_{[\mu}{}^{[a} T_{\nu\alpha]}{}^{b]} + \frac{2mM}{(d-1)(d-2)}
e_{[\mu}{}^a e_\nu{}^b \Phi_{\alpha]} \nonumber \\
D_{[\mu} T_{\nu\alpha]}{}^a &=& - {\cal F}_{[\mu\nu,\alpha]}{}^a -
\frac{2m}{(d-2)} e_{[\mu}{}^a {\cal B}_{\nu\alpha]} \nonumber \\
D_{[\mu} {\cal B}_{\nu]}{}^{ab} &=& m {\cal F}_{\mu\nu}{}^{ab} +
\frac{M}{(d-1)} e_{[\mu}{}^{[a} \Pi_{\nu]}{}^{b]} \nonumber \\
D_{[\mu} {\cal B}_{\nu\alpha]} &=& - {\cal B}_{[\mu,\nu\alpha]} -
\frac{m}{2} T_{[\mu\nu,\alpha]} \\
D_{[\mu} \Pi_{\nu]}{}^a &=& \frac{(d-1)}{(d-2)} M {\cal
B}_{[\mu,\nu]}{}^a + \frac{(d-1)}{(d-2)} mM T_{\mu\nu}{}^a +
\frac{m^2}{(d-2)} e_{[\mu}{}^a \Phi_{\nu]} \nonumber \\
D_{[\mu} \Phi_{\nu]} &=& - \Pi_{[\mu,\nu]} - \frac{2(d-1)}{(d-2)} M
{\cal B}_{\mu\nu} \nonumber
\end{eqnarray}

Let us cinsider the most general ansatz for the Lagrangian in terms of
these objects:
\begin{eqnarray}
{\cal L}_0 &=& a_1 \epfour {\cal F}_{\mu\nu}{}^{ab}
{\cal F}_{\alpha\beta}{}^{cd} + a_2 \eptwo {\cal B}_\mu{}^{ac} 
{\cal B}_\nu{}^{bc} + a_3 \eptwo \Pi_\mu{}^a \Pi_\nu{}^b \nonumber \\
&& + a_4 \epthree {\cal F}_{\mu\nu}{}^{ab} \Pi_\alpha{}^c + a_5
\epthree T_{\mu\nu}{}^a {\cal B}_\alpha{}^{bc} + a_6 \eptwo
{\cal B}_\mu{}^{ab} \Phi_\nu
\end{eqnarray}
Not all these terms are independent as can be shown by considering the
following identities valid up to the total derivative:
$$
\epfour D_\mu [ {\cal F}_{\nu\alpha}{}^{ab} {\cal B}_\beta{}^{cd} ]
\approx 0, \qquad
\epthree D_\mu [ {\cal B}_\nu{}^{ab} \Pi_\alpha{}^c ] \approx 0
$$
Using the differential identities given above it is rather
straightforward task to obtain two relations on these six terms:
$$
\frac{m}{2} X_1 + \frac{8(d-3)m}{(d-2)} X_2 + \frac{2(d-3)M}{(d-1)}
X_4 + \frac{2(d-3)M^2}{(d-2)} X_5 - \frac{4(d-3)mM}{(d-1)} X_6 = 0
$$
$$
- \frac{2(d-1)M}{(d-2)} X_2 + \frac{2(d-2)M}{(d-1)} X_3 + \frac{m}{2}
X_4 - \frac{(d-1)mM}{2(d-2)} X_5 + m^2 X_6 = 0
$$
where $X_i$ denotes term with the coefficient $a_i$. Thus we have a
whole family of possible solutions with two free parameters. Let us
provide here just two concrete examples. The first one is the choice
of the authors of \cite{PV10}:
\begin{equation}
{\cal L} = \frac{(d-2)}{32(d-3)M^2} \epfour {\cal F}_{\mu\nu}{}^{ab}
{\cal F}_{\alpha\beta}{}^{cd} + \frac{1}{2M^2} \eptwo 
{\cal B}_\mu{}^{ac} {\cal B}_\nu{}^{bc} - \frac{(d-2)}{2(d-1)M}
\eptwo {\cal B}_\mu{}^{ab} \Phi_\nu
\end{equation}
This solution admits the massless limit for the non-zero cosmological
constant (but not the partially massless one) and is applicable for $d
\ge 4$ only. The other possible solution has the form:
\begin{equation}
{\cal L} = - \frac{1}{2m^2} \eptwo {\cal B}_\mu{}^{ac} 
{\cal B}_\nu{}^{bc} + \frac{(d-2)^2}{2(d-1)^2m^2} \eptwo \Pi_\mu{}^a
\Pi_\nu{}^b - \frac{1}{4m} \epthree T_{\mu\nu}{}^a 
{\cal B}_\alpha{}^{bc}
\end{equation}
This solution does not admit the massless limit (independently of the
cosmological constant) but it is nicely works for $d=3$ case. Moreover
the structure of the Lagrangian is similar to the three-dimensional
ones considered in this work.

\end{document}